\documentclass[doublecol]{epl2} 

\usepackage{amsmath,amssymb}

\title{Effects of local curvature on epithelia tissue - coordinated rotational movement and other spatiotemporal arrangements}
\shorttitle{Effects of local curvature on epithelia tissue} 

\author{L. Happel\inst{1} \and D. Wenzel\inst{1} \and A. Voigt\inst{1,2,3}}
\shortauthor{L. Happel \etal}

\institute{                    
  \inst{1} Institute of Scientific Computing, Technische Universit\"at Dresden - 01062 Dresden, Germany\\
  \inst{2} Centre for Systems Biology Dresden (CSBD) - Pfotenhauerstr. 108, 01307 Dresden, Germany\\
  \inst{3} Cluster of Excellence - Physics of Life, Technische Universit\"at Dresden - 01062 Dresden, Germany
}

\abstract{
Coordinated movements of epithelia tissue are linked with active matter processes. We here consider the influence of curvature on the spatiotemporal arrangements and the shapes of the cells. The cells are represented by a multiphase field model which is defined on the surface of a sphere. Besides the classical solid and liquid phases, which depend on the curvature of the sphere, on mechanical properties of the cells and the strength of activity, we identify a phase of global rotation. This rotation provides a coordinated cellular movement which can be linked to tissue morphogenesis. This investigation on a sphere is a first step to investigate the delicate interplay between topological constraints, geometric properties and collective motion. Besides the rotational state we also analyse positional defects, identify global nematic order and study the associated orientational defects.}

\begin{document}

\maketitle

\section{Introduction}

Epithelia tissues display a large variety of forms, from flat to highly curved. They are involved in morphogenetic events leading to complex multicellular organisms and are intensively studied in-vivo, in-vitro and in-silico, see \cite{Friedl_NRMCB_2009,Rorth_ARCDB_2009,Scarpa_JCB_2016,Hakim_RPP_2017,Alert_ARCMP_2020} for reviews. While the mechanics of epithelia tissues are well understood in flat space and also various properties have been identified which trigger out-of-plane deformations to start the development towards highly curved structures \cite{Saw_N_2017,Maroudas-Sacks_NP_2020}, less is known about the influence of curvature on the spatiotemporal arrangements and shapes of the cells. A physical framework which accounts for the curvature of epithelia would therefore be highly desirable to better understand the development of multicellular organisms.

The influence of curvature on the motility of epithelial cells has been studied recently using Madin-Darby canine kidney (MDCK) cells \cite{Yevick_PNAS_2015,Xi_NC_2017,Luciano_NP_2021}. The considered geometries are tubes, rods or wavy patterns. They all have in common a non-zero mean but zero Gaussian curvature. Furthermore topological constraints resulting from a closed surface, as e.g. an egg chamber in early embryogenesis, are not present in these investigations. These additional mathematical effects are accounted for in epithelial acini \cite{Tanner_PNAS_2012,Wang_PNAS_2013}, which are spherical like structure with the epithelium surrounding a lumen. These experiments demonstrate not only an adjustment of cell shapes to curvature but also coordinated rotational movement. Similar persistent and continuous azimuthal rotation are observed during Drosophila oogenesis \cite{Haigo_Science_2011,Cetera_NC_2014}.

We will consider an idealized spherical surface to acquire deeper insights into these complex phenomena of active matter, tissue mechanics, and geometrical and topological constraints. Various modeling approaches have been considered to account for the multiscale aspects involved in the mechanics of epithelia tissue. 

On a coarse-grained continuous description active gel models have shown to reproduce significant mechanical features, see e.g. \cite{Doostmohammadi_NC_2018} for a review. First attempts to formulate and simulate these models on surfaces also exist \cite{Nestler_JNS_2017,Nitschke_PRSA_2018,Pearce_PRL_2019,nitschke2019hydrodynamic,Mietke_PNAS_2019,Mietke_PRL_2019,Nitschke_PRSA_2020}. However, most of these approaches are restricted to axisymmetric surfaces or intrinsic coupling terms with the geometry. For the effect of additional extrinsic coupling terms we refer to \cite{Napoli_PRL_2012,Nitschke_PRSA_2020,Nestler_SM_2020,Nestler_arXiv_2021}. Furthermore, these coarse-grained models do not allow to analyse the spatiotemporal arrangements and shapes of the cells. 

Another class are vertex models. For flat geometries they are a versatile tool to reproduce e.g. solid-liquid phase transitions, number of neighbor distributions and stress fields in epithelia tissue \cite{Nagai_PMB_2001,Staple_EPJE_2010,Fletcher_BPJ_2014,Li_BPJ_2014,Bi_PRX_2016}. Various 3D approaches have been developed and also extensions towards formulations on curved surfaces \cite{Sussman_PRR_2020}, where the effect of curvature on the solid-liquid transition is analysed on spherical surfaces. 

We here consider a multiphase field approach, see \cite{Nonomura_PLOS_2012,Camley_PNAS_2014,Loeber_SR_2015,Palmieri_SR_2015,Mueller_PRL_2019,Wenzel_JCP_2019,Loewe_PRL_2020,Wenzel_arXiv_2021}, which allows for cell deformations and detailed cell-cell interactions, as well as subcellular details to resolve the mechanochemical interactions underlying cell migration. Also topological changes, such as T1 transitions, follow naturally in this framework. Multiphase field models, together with efficient numerics and appropriate computing power, allow to analyse the connection of single cell behaviour with collective behaviour, at least for moderate numbers of cells. They already led to quantitative predictions of many generic features in multicellular systems \cite{Peyret_BJ_2019,balasubramaniamEtAl2021,Wenzel_arXiv_2021}. We here formulate a multiphase field model on the surface of a sphere. We analyse the solid-liquid transition and identify a rotational mode between the solid and the liquid phase. We further consider the cell shapes, their neighbor relations and emerging nematic liquid crystal order.

\section{Modeling}

We consider two-dimensional phase field variables $\phi_i(\mathbf{x},t)$, one for each cell. Values of $\phi_i = 1$ and $\phi_i = -1$ denote the interior and the exterior of a cell, respectively. The cell boundary is defined implicitly by the region $-1 < \phi_i <  1$. The dynamics for each $\phi_i$ is considered as
\begin{equation}
    \partial_t \phi_i + v_0 (\mathbf{v}_i \cdot \nabla_{\cal{S}} \phi_i) = \Delta_{\cal{S}} \frac{\delta \mathcal{F} }{\delta \phi_i}, 
    \label{eq:phi}
\end{equation}
for $i = 1, \ldots, N$, where $N$ denotes the number of cells. $\mathcal{F}$ is a free energy and $\mathbf{v}_i$ is a vector field used to incorporate active components, with a self-propulsion strength $v_0$. The operators $\nabla_{\cal{S}}$ and $\Delta_{\cal{S}}$ denote the surface divergence and Laplace-Beltrami operator on the sphere $\cal{S}$, respectively. All quantities are non-dimensional quantities. As in previous studies \cite{Marth_JRSI_2015,Marth_IF_2016,Wenzel_JCP_2019,Wenzel_CMAM_2021,Wenzel_arXiv_2021,Jain_arXiv_2021} we consider conserved dynamics. The free energy $\mathcal{F} = \mathcal{F}_{CH} + \mathcal{F}_{INT}$ contains passive contributions, where
\begin{eqnarray}
\mathcal{F}_{CH} &=& \sum_{i=1}^N \frac{1}{Ca}\int_{\cal{S}} \frac{\epsilon}{2}\|\nabla_{\cal{S}} \phi_i\|^2 + \frac{1}{\epsilon}W(\phi_i)\,\text{d}{\cal{S}}, \\
\label{eq:IntEnergy}
\mathcal{F}_{INT} &=& \sum_{i=1}^N \frac{1}{In}\int_{\cal{S}} B(\phi_i) \sum_{j\neq i} w(\phi_j)\,\text{d}{\cal{S}}, 
\end{eqnarray}
with non-dimensional capillary and interaction number, $Ca$ and $In$, respectively. The first is a Cahn-Hilliard energy, with $W(\phi_i) = \frac{1}{4}(\phi_i^2 - 1)^2$ a double-well potential and $\epsilon$ a small parameter determining the width of the diffuse interface. Due to this energy non-interacting cells tend to become circular. The second is an interaction energy with $B(\phi_i) = \frac{\phi_i + 1}{2}$, a simple shift of $\phi_i$ to values in $[0,1]$ and a cell-cell interaction potential 
\begin{equation}
    w(\phi_j) = 1 - (a+1)\left(\frac{\phi_j -1}{2}\right)^2 + a \left(\frac{\phi_j -1}{2}\right)^4
    \label{eq:interaction}
\end{equation}
approximating a short range potential, which is only active within the interior of the cell and its diffuse boundary. The approach offers the possibility to consider repulsive as well as attractive interactions which can be tuned by parameter $a$, see \cite{Jain_arXiv_2021}. We here consider $a=2$ which models repulsive and attractive interactions. For the definition of $\mathbf{v}_i$ in eq. (\ref{eq:phi}), we follow the simplest possible approach, which can be considered as a generalization of active Brownian particles \cite{Fily_PRL_2012,Redner_PRE_2013,Wysocki_EPL_2014} to deformable objects \cite{Loewe_PRL_2020}. In this approach the specified propulsion speed $v_0$ is the same for each cell, but the specified direction of motion, determined by the angle $\theta_i$ is controlled by rotational noise $d \theta_i(t) = \sqrt{2 D_r} d W_i(t)$, with diffusivity $D_r$ and a Wiener process $W_i$, which results in $\mathbf{v}_i = (\cos{\theta_i}, \sin{\theta_i})$, within a local coordinate system for the tangent plane at the centre of mass of cell $i$. For consistency we define one of the orthonormal vectors with respect to the projected velocity vector of the previous time step.  

\section{Numerical issues}

The resulting system is solved by surface finite elements \cite{dziuk2013finite,nestler2019finite} within the toolbox AMDiS \cite{Vey_CVS_2007,Witkowski_ACM_2015}. A piecewise flat surface triangulation is used which is adaptively refined within the diffuse interface. The other approximations in space and time follow the validated strategies for the flat case considered in \cite{Wenzel_arXiv_2021}. Following \cite{Salvalaglio_MMAS_2021} we include a degenerate term in the Cahn-Hilliard energy. This modification does not affect the asymptotic behaviour as $\epsilon \to 0$ but helps to ensure $\phi_i \in [-1,1]$ and to increase accuracy, see \cite{Salvalaglio_MMAS_2021,Backofen_IJNAM_2019}. To efficiently solve the resulting system of equations each cell is considered on a different core. Due to the short-range interaction communication can be reduced and the implementation essentially scales with the number of cells \cite{Praetorius_NIC_2017}. In order to achieve a large area fraction the initial condition is constructed by regularly arranging cells with an equilibrium $\tanh$-profile. All cells are of equal size. All simulations correspond to an area fraction of $94\%$. The considered parameters are comparable to those considered in the plane. We essentially use $\epsilon = 0.006$, $In = 6.0$ and vary $Ca \in [150,450]$ and $v_0 \in [0.0, 3.0]$. We consider $N = {6, 12, 32, 92}$ and adapt the radius of the sphere to allow for a constant cell size. For $N = 32$ the radius of the sphere is $1$. 

\section{Results}

\begin{figure}[htb]
    \centering
    \includegraphics[width=0.48\textwidth]{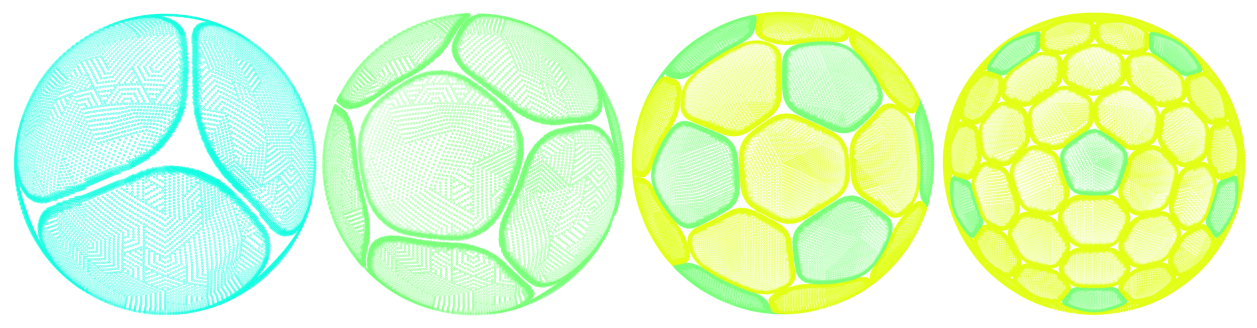} 
    \caption{Equilibrium configuration for $6$, $12$, $32$ and $92$ cells. The color coding shows the number of neighbors (blue - four, green - five, yellow - six). The radius of the sphere is scaled to be equal. The fine details show the adaptive refined mesh for each cell.}
    \label{fig1}
\end{figure}

In order to understand the active system we first consider the passive case with $v_0 = 0$ and ask for the equilibrium configuration. In contrast to the flat situation, where this is given by a hexagonal ordering the topology of the sphere introduces defects in the ground state. The optimal arrangement of the cells is related to the Thomson problem \cite{Thomson_PM_1904} and various numerical methods have been used to find the minimal energy configuration for different $N$. While this remains a difficult task for large $N$, for the considered cases the configurations are known and have also been found using related phase field crystal models \cite{Backofen_PRE_2010,Backofen_MMS_2011}. Figure \ref{fig1} shows the corresponding cell configurations, with an appropriate color coding indicating the number of neighbors. 
These results correspond to $Ca = 150$. Varying $Ca$ does not influence the arrangements, it only modifies the shape of the cells. The equilibrium configuration can be used to define a standard deviation of the number of neighbors $std_{eq}$. For $N = 6$, which corresponds to the arrangement of a cube and $N = 12$, corresponding to a dodecahedron, we have $std_{eq} = 0.0$ as all cells have 4 or 5 neighbors, respectively. For $N=32$ and $92$, there are 12 cells with 5 neighbors, and 20 and 80 cells with 6 neighbors, corresponding to $std_{eq} = 0.47$ and $0.34$, respectively. A threshold on the deviation from these equilibrium values $d = \|std - std_{eq}\| \leq 0.06$ is used to identify the solid and the liquid state. This criteria is related to the coordination number in flat space, which is frequently used as an easily accessible structural property to identify solid-liquid transitions. We here consider a second criteria the mean square displacement ($msd$) of the centres of mass of all cells as used in \cite{Loewe_PRL_2020}. Before the calculation of the $msd$ the radius of the sphere is scaled to be equal for the different numbers of cells $N$. We again identify a threshold $msd \leq 0.02$ to distinguish between solid and liquid. As known from flat space both activity $v_0$ and cell deformability, here considered through $Ca$, can alter the solid-liquid transition. On the sphere we have some additional degrees of freedom which can be identified by combining the two criteria. 
\begin{figure}[htb]
    \centering
    \includegraphics[width=0.48\textwidth]{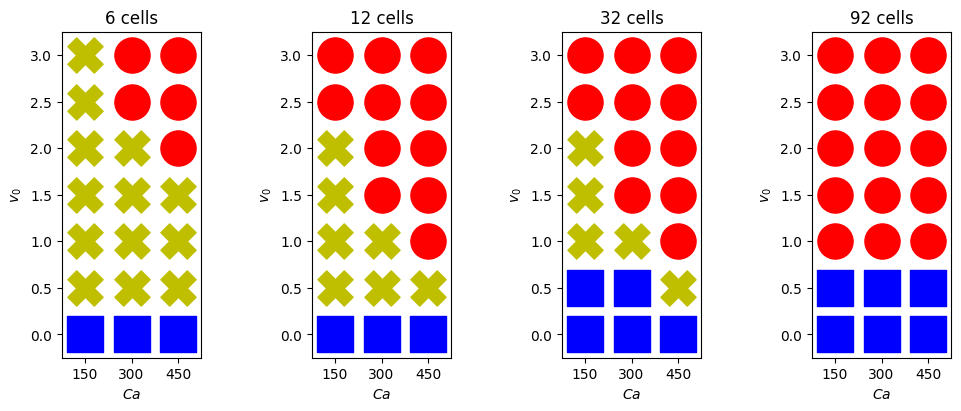} 
    \caption{Phase diagram for different numbers of cells. Liquid states are marked with red circles, rotation with a yellow cross and solid with a blue square.}
    \label{fig2}
\end{figure}
We end up with four cases: If both $msd$ and $d$ are below their thresholds, the cells remain in the stable configuration and the system is not moving as a whole. The system is in the solid state. If both quantities are above their thresholds, the cells are moving and are out of the stable configuration. This indicates a liquid state. If $msd$ is above, but $d$ below the threshold, the cells remain in the stable configuration but the system moves. This indicates coordinated rotational movement. We call this rotation state. The fourth possibility was never observed in our studies. Figure \ref{fig2} shows the resulting phase diagram for the considered parameters. 
\begin{figure}[htb]
    \centering
    \includegraphics[width=0.48\textwidth]{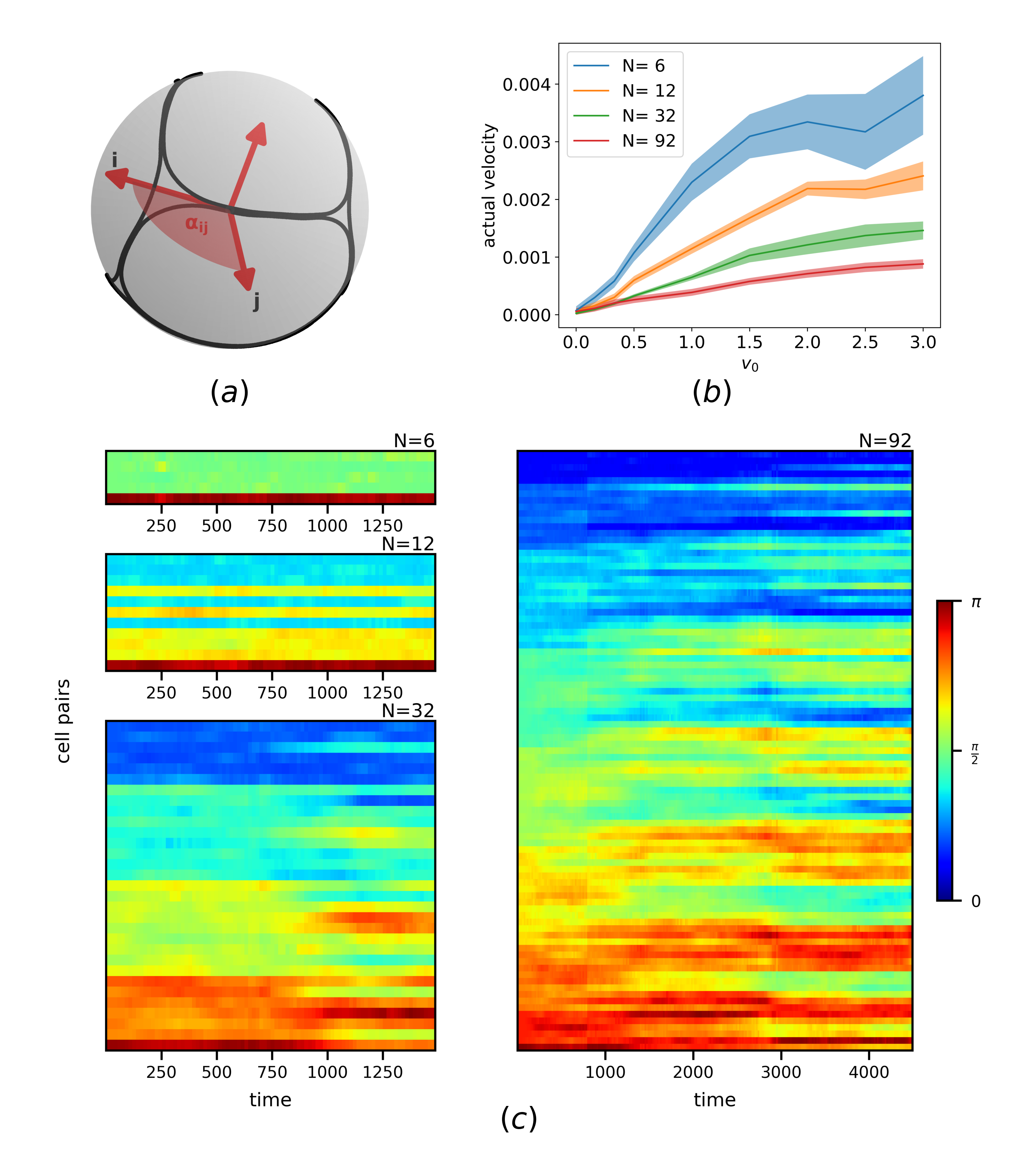} 
    \caption{Rotating stable configuration for $N=6, 12$ and $32$ identified through constant angles in kymograph. For $N=92$ this configuration is not present. (a) Schematic picture to identify the angle for cell pairs, (b) Mean absolute velocity of cells as function of $v_0$, averaged over all time steps and several simulations. Before the calculation of $v_0$ the radius of the sphere is scaled to be equal. The shaded regions show the variance of the velocity, and (c) kymographs for 6, 12, 32, and 92 cells, each row corresponds to one cell pair. For clarity we only show pairs which include cell $0$.}
    \label{fig3}
\end{figure}
To further confirm the identified global rotation we plot the pairwise angles between the centres of mass of all cells over time, see Fig. \ref{fig3}. They are exemplary shown for $v_0 = 1.5$ and $Ca = 150$ and one cell fixed.

\begin{figure}[htb]
    \centering
    \includegraphics[width=0.48\textwidth]{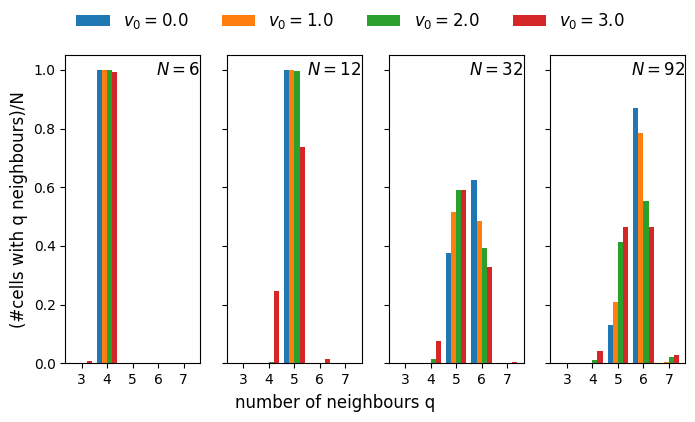} 
    \caption{Number of neighbor distribution for various $v_0$ and $Ca = 150$.}
    \label{fig4}
\end{figure}

As these angles are characteristic for the equilibrium configuration the constant values in the kymographs confirm a rotation if $msd$ is above the considered threshold. At least for low numbers of cells, $N = 6$ and  $12$ this is the case and the rotation state is characteristic for the phase diagram. Due to the random selection of the direction of movement in our model, the rotation is not persistent, but randomly changes its axis. This randomness is also responsible for the disappearance of a rotation state for larger numbers of cells. For $N = 32$ the angles remain only constant over some time period, which confirms the presence of a rotation state also in this case. However, it is less stable. For $N = 92$ the angles change, which is consistent with the identified liquid state in the phase diagram. Other modeling approaches with a more deterministic relation of the direction to the cell shape or internal degrees of freedom, see \cite{Mueller_PRL_2019,Wenzel_arXiv_2021} for realisations in flat space, will probably cure this issue and lead to global persistent rotation also for larger numbers of cells, similar to the rotation observed for active crystals on a sphere using a related active phase field crystal model \cite{Praetorius_PRE_2018}. Another property common for many active systems is also shown in Fig. \ref{fig3}. We plot the mean absolute velocity as a function of $v_0$. Again this is only shown for $Ca = 150$. The average velocity is computed from the position of the centres of mass at each time step. The shaded region shows the variance. The deviation from the expected increase with $v_0$ corresponds to the transition from rotation to less ordered movement in the liquid phase.   

\begin{figure}[htb]
    \centering
    \includegraphics[width=0.48\textwidth]{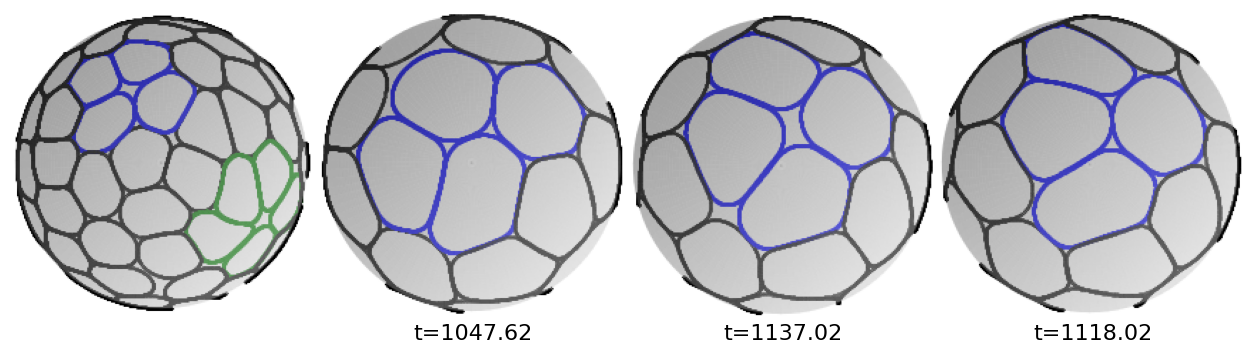} 
    \caption{Example for two rosettes, which are color coded in simulation with 92 cells and a time series of a T1 transition with 32 cells. The parameters are $Ca = 150$ and $v_0 = 3.0$.}
    \label{fig5}
\end{figure}

We are next interested in the distribution of the number of neighbors as a function of $v_0$. We compute this quantity in each time step and average over all times. Fig. \ref{fig4} shows the resulting distributions. While the deviation for $N=6$ and $12$ are minor, for $32$ and $92$ cells we observe a clear broadening of the distribution and a shift of the maximal value towards $5$. The increased heterogeneity of the numbers of neighbors for increased $v_0$ is also visible by looking at snapshots. Fig. \ref{fig5} shows configurations with rosettes and also a sequence for a T1 transition. As in flat space these are natural features of the multiphase field model and their occurrence increases with $v_0$.

\begin{figure}[htb]
    \centering
    \includegraphics[width=0.38\textwidth]{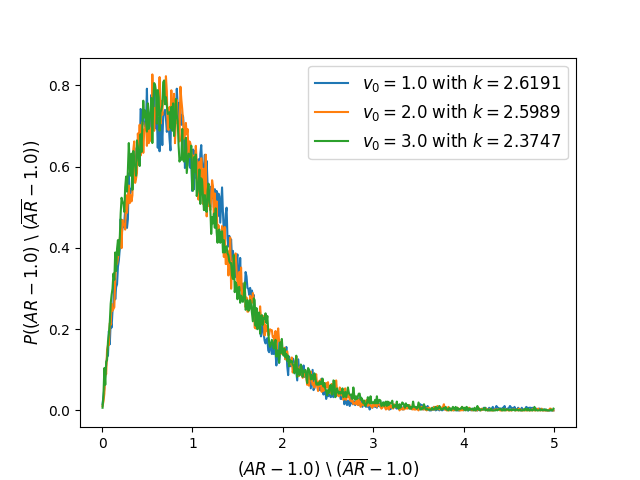} 
    \caption{Distribution of aspect ratio for 92 cells with $Ca = 150$ and various $v_0$, together with a best fit for $k$.}
    \label{fig6}
\end{figure}

While in the equilibrium state the cells are isotropic, with no preferred orientation, this changes in the liquid phase. The snapshots in Fig. \ref{fig5} already indicate a deformation of the cells. We quantify this by computing the aspect ratio $AR$, as the ratio of the major and the minor axis of an ellipse fitted to the cell. As the exact calculation on a surface is tedious we only use an approximation assuming the surface to be locally flat in the centre of mass of each cell. This is valid as long as the cell size is small compared to the surface of the sphere, which is the case for $N=92$. We compute $x = (AR - 1.0) / (\overline{AR} - 1.0$) with $\overline{AR}$ the mean of the aspect ratio of the cells. In \cite{Atia_NP_2018} it is shown that $x$ is universal and can be described by a $k$-Gamma distribution $pdf(x,k) = k^k x^{k-1} e^{-kx} / \Gamma(k$), with Legendre-Gamma function $\Gamma$. Fig. \ref{fig6} shows the obtained results for 92 cells with $Ca = 150$ and various values for $v_0$.    
According to \cite{Atia_NP_2018} diverse epithelia systems, including Madin-Darby canine kidney (MDCK) cells, Human bronchial epithelia cells (HBECs) and the Drosophila embryo during ventral furrow formation, show a distribution with k in a narrow range between 2 and 3. The computed values for $k$ in Fig. \ref{fig6} are within this range. 

\begin{figure}[htb]
    \centering
    \includegraphics[width=0.48\textwidth]{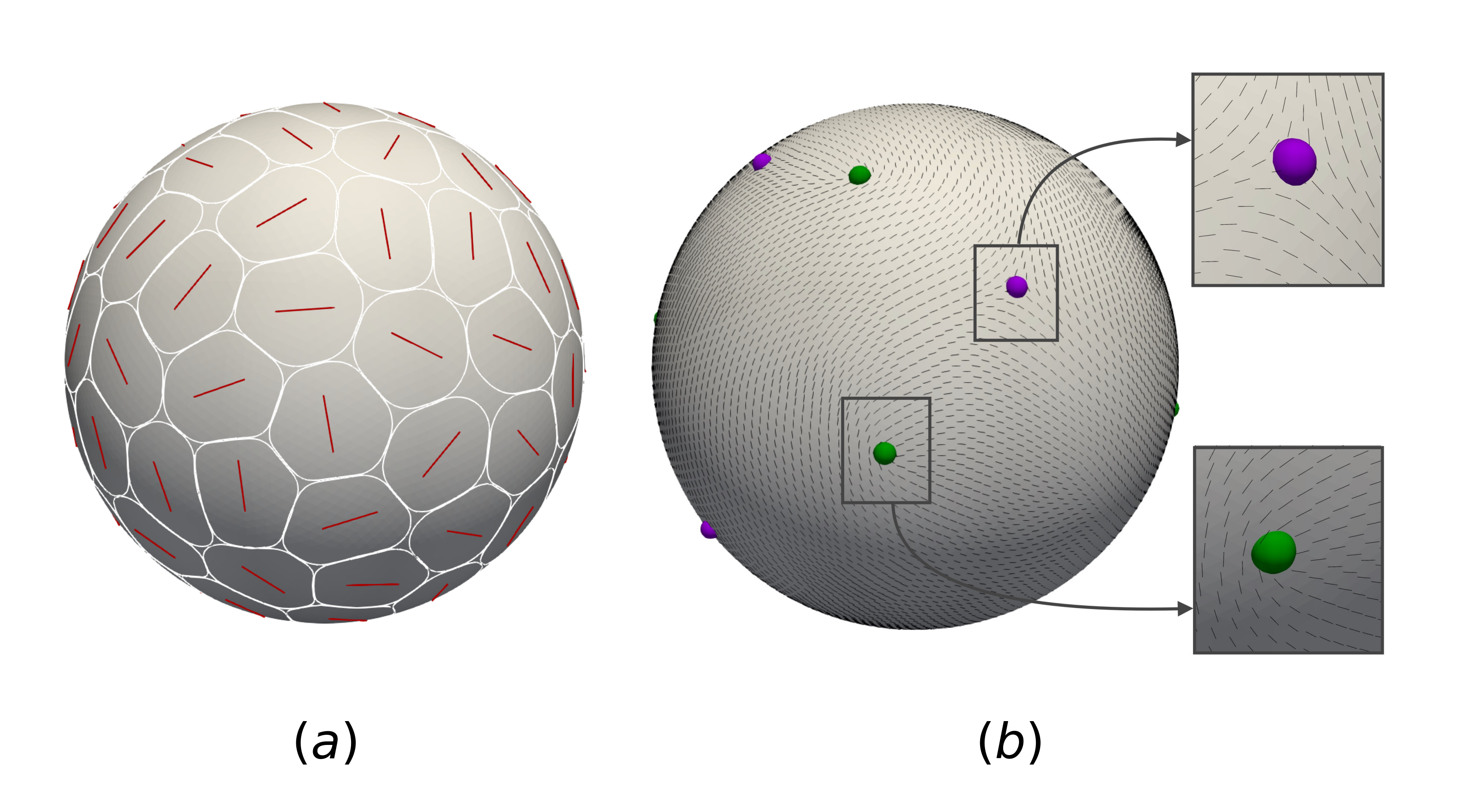} 
    \caption{(a) Major axis of each cell. (b) Nematic director field together with topological defects ($+1/2$ - green, $-1/2$ - purple), see inlet for closeup of director field. The coarse-grained continuous description is obtained after appropriate averaging and interpolation. (a) and (b) show the same time instance and same perspective for a simulation with 92 cells. }
    \label{fig7}
\end{figure}

This agreement further indicates the possibility to define a nematic order induced by the cell shapes. Taking the major axis in the centre of mass of each cell as a discrete orientated object allows to define a coarse-grained tangential director field. This induces nematic order and leads the presence of topological defects, see Fig. \ref{fig7}. 
This coarse-graining allows to establish a connection to surface active gel models, similar as considered for the flat case, where epithelia tissue are successively described by these theories \cite{Doostmohammadi_NC_2018}. We compute an averaged director field by averaging the major axis in each cell with their neighboring cells. This is done within the tangent planes of the centres of mass of each cell. We interpolate between the resulting directors using a Gaussian kernel where the standard deviation corresponds to the radius of one cell. This defines a global tangential nematic order on the surface. From this director field (and the corresponding orthogonal vector-field in the tangent plane) we compute a global tangential $Q$-Tensor field $\boldsymbol{q}$. Defects in this field are identified by maxima of $||\nabla \boldsymbol{q} ||^2$. To determine the topological charge we project $\boldsymbol{q}$ in the tangent plane around the defect and transform this into the $xy$- plane. This allows to use established methods in flat space for this issue. Following a physical approach developed in \cite{Hesselink_PV_1994}, which has been successfully used for multiphase field models in \cite{Wenzel_CMAM_2021, Wenzel_arXiv_2021}, we use the sign of $\delta=\frac{\partial q_{11}}{\partial x}\frac{\partial q_{12}}{\partial y}-\frac{\partial q_{11}}{\partial y}\frac{\partial q_{12}}{\partial x}$ to distinguish between $+1/2$ and $-1/2$ defects where $q_{11}$ and $q_{12}$ denote the two independent entries of a $2$-dimensional $Q$-tensor.  We track the trajectories of the defects using the software \textit{trackpy 0.5.0} \cite{Allan_2021} which implements the Crocker-Grier linking algorithm \cite{Crocker_JoCaIS_1996}. The resulting data can now be compared with results of surface active gel models on a sphere, see \cite{ZhangDesernoTu_2020,NestlerVoigt_2022}, and other modeling approaches and experimental data.

\begin{figure}[htb]
    \centering
    \includegraphics[width=0.48\textwidth]{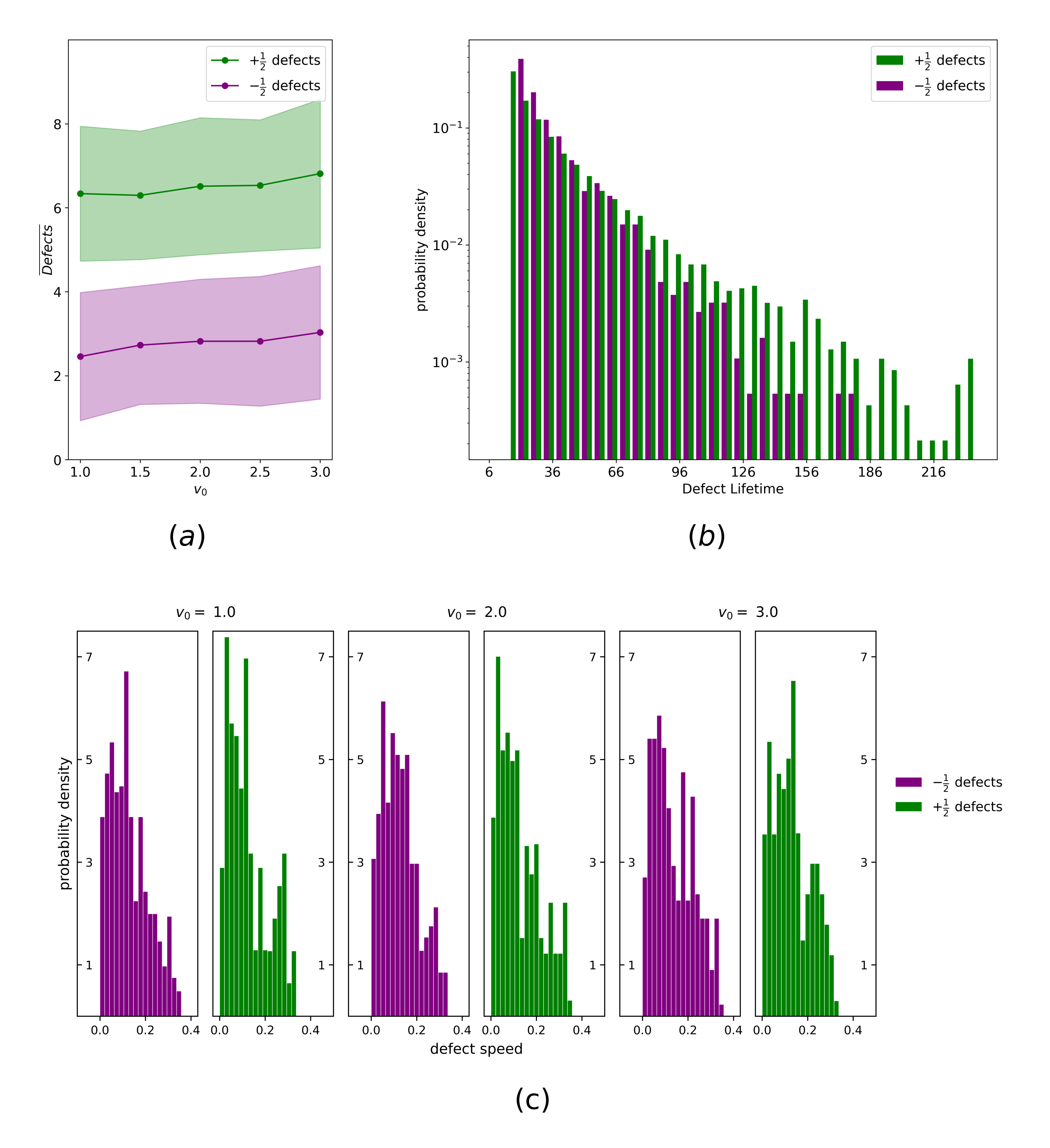} 
    \caption{(a) Average number of defects together with standard deviation for different activities $v_0$ and $Ca=300.0$. (b) Distribution of life time of defects. (c) Defect velocities for different $v_0$ and $Ca=300.0$. }
  \label{fig8}
\end{figure}

First, due to topological reasons the sum of the topological charges of all defects has to be equal to the Euler characteristic of the surface, which is $2$ for the considered sphere. This is at least on average fulfilled, see Fig. \ref{fig8}(a). It shows the number of defects averaged over all time steps and for various $v_0$. On average we have roughly six $+1/2$ defects and two $-1/2$ defects, the numbers slightly increase with $v_0$, which gives the desired topological charge. The deviation from this average values also indicate a frequent creation and annihilation of defect pairs. The presence of $-1/2$ defects and the fluctuation in the overall number qualitatively differs from results for surface active gels models and also known experimental results. A prominent example is \cite{Keberetal_2014}, considering microtubles and molecular motors encapsulated within a spherical lipid vescicle. In this setting the topological constraint suppresses the creation of additional defects and the systems remains close to the equilibrium configuration with four $+1/2$ defects. In the active setting they oscillate. The behaviour is reproduced by various simulation approaches \cite{Keberetal_2014,Alaimoetal_2017,ZhangDesernoTu_2020,NestlerVoigt_2022}. At least the continuous approaches \cite{ZhangDesernoTu_2020,NestlerVoigt_2022} also demonstrate that regular defect oscillation are only present for low activities. If the activity is increased they become irregular and in principle also the creating and annihilation of defects pairs becomes possible. Experimental data for such defects in spherical epithelia tissue are not know. Fig. \ref{fig8}(b) analyses the life time of the defects, which indicates a stronger persistence of $+1/2$ defects and Fig. \ref{fig8}(c) compares the average velocity between $+1/2$ and $-1/2$ defects for different $v_0$.  
In agreement with results in flat space \cite{Wenzel_arXiv_2021} no significant difference can be seen for the velocity. As expected the distributions shift towards larger velocities for increasing $v_0$. However, also this effect is present for both $+1/2$ and $-1/2$ defects.

\section{Conclusion}

We have considered a minimal multiphase field model for epithelia tissue on the surface of a sphere. The topological constraint induces positional defects in the equilibrium configuration. Deviations from this configuration and cell movements have been used to identify a global rotating state between the solid and the liquid state. Even for the considered random propulsion mechanism this rotation state is persistent, at least for small cell numbers. Further investigations for larger cell numbers in the liquid state indicate changes in the number of neighbor distribution and shape deformations. The elongation of cells is used to define a global nematic order. This allows to analyse the emerging orientational defects. They again follow the required topological constraint but in contrast with other systems and coarse-grained computational results obtained with surface active gel models the topological constraint does not suppress the creation and annihilation of additional defects. This might indicate fundamental differences between these systems and asks for experimental investigation on spherical epithelia tissue.  

\acknowledgments
A.V. acknowledges support by the German Research Foundation (DFG) under Grant FOR3013 and Vo899/19-2. We further acknowledge computing resources provided at J\"ulich Supercomputing Center under Grant No. PFAMDIS and at ZIH at TU Dresden. \\

\bibliography{library}

\begin{thebibliography}{10}
\expandafter\ifx\csname url\endcsname\relax\def\url#1{\texttt{#1}}\fi

\bibitem{Friedl_NRMCB_2009}
\Name{Friedl P. \and Gilmour D.} \REVIEW{Nature Rev. Mol. Cell
  Biol.}{10}{2009}{445}.

\bibitem{Rorth_ARCDB_2009}
\Name{Rorth P.} \REVIEW{Ann. Rev. Cell Devel. Biol.}{25}{2009}{407}.

\bibitem{Scarpa_JCB_2016}
\Name{Scarpa E. \and Mayor R.} \REVIEW{J. Cell Biol.}{212}{2016}{143}.

\bibitem{Hakim_RPP_2017}
\Name{Hakim V. \and Silberzan P.} \REVIEW{Rep. Prog. Phys.}{80}{2017}{076601}.

\bibitem{Alert_ARCMP_2020}
\Name{Alert R. \and Trepat X.} \REVIEW{Ann. Rev. Cond. Matt.
  Phys.}{11}{2020}{77}.

\bibitem{Saw_N_2017}
\Name{Saw T., Doostmohammadi A., Nier V., Kocgozlu L., Thampi S., Toyama Y.,
  Marcq P., Lim C., Yeomans J. \and Ladoux B.} \REVIEW{Nature}{544}{2017}{212}.

\bibitem{Maroudas-Sacks_NP_2020}
\Name{Maroudas-Sacks Y., Garion L., Shani-Zerbib L., Livshits A., Braun E. \and
  Keren K.} \REVIEW{Nat. Phys.}{17}{2021}{251}.

\bibitem{Yevick_PNAS_2015}
\Name{Yevick H.~G., Duclos G., Bonnet I. \and Silberzan P.} \REVIEW{Proc. Nat.
  Acad. Sci. (USA)}{112}{2015}{5944}.

\bibitem{Xi_NC_2017}
\Name{Xi W., Sonam S., Saw T.~B., Ladoux B. \and Lim C.~T.} \REVIEW{Nat.
  Comm.}{8}{2017}{1517}.

\bibitem{Luciano_NP_2021}
\Name{Luciano M., Xue S.-L., De~Vos W.~H., Redondo-Morata L., Surin M., Lafont
  F., Hannezo E. \and Gabriele S.} \REVIEW{Nat. Phys.}{17}{2021}{1382}.

\bibitem{Tanner_PNAS_2012}
\Name{Tanner K., Mori H., Mroue R., Bruni-Cardoso A. \and Bissell M.~J.}
  \REVIEW{Proc. Nat. Acad. Sci. (USA)}{109}{2012}{1973}.

\bibitem{Wang_PNAS_2013}
\Name{Wang H., Lacoche S., Huang L., Xue B. \and Muthuswamy S.~K.}
  \REVIEW{Proc. Nat. Acad. Sci. (USA)}{110}{2013}{163}.

\bibitem{Haigo_Science_2011}
\Name{Haigo S.~L. \and Bilder D.} \REVIEW{Science}{331}{2011}{1071}.

\bibitem{Cetera_NC_2014}
\Name{Cetera M., Juan G. R. R.-S., Oakes P.~W., Lewellyn L., Fairchild M.~J.,
  Tanentzapf G., Gardel M.~L. \and Horne-Badovinac S.} \REVIEW{Nat.
  Comm.}{5}{2014}{5511}.

\bibitem{Doostmohammadi_NC_2018}
\Name{Doostmohammadi A., Ignes-Mullol J., Yeomans J. \and Sagues F.}
  \REVIEW{Nat. Comm.}{9}{2018}{3246}.

\bibitem{Nestler_JNS_2017}
\Name{Nestler M., Nitschke I., Praetorius S. \and Voigt A.} \REVIEW{J. Nonlin.
  Sci.}{28}{2018}{147}.

\bibitem{Nitschke_PRSA_2018}
\Name{Nitschke I., Nestler M., Praetorius S., L\"owen H. \and Voigt A.}
  \REVIEW{{Proc. Roy. Soc. A}}{474}{2018}{20170686}.

\bibitem{Pearce_PRL_2019}
\Name{Pearce D. J.~G., Ellis P.~W., Fernandez-Nieves A. \and Giomi L.}
  \REVIEW{Phys. Rev. Lett.}{122}{2019}{168002}.

\bibitem{nitschke2019hydrodynamic}
\Name{Nitschke I., Reuther S. \and Voigt A.} \REVIEW{Phys. Rev.
  Fluids}{4}{2019}{044002}.

\bibitem{Mietke_PNAS_2019}
\Name{Mietke A., Juelicher F. \and Sbalzarini I.~F.} \REVIEW{Proc. Nat. Acad.
  Sci. (USA)}{116}{2019}{29}.

\bibitem{Mietke_PRL_2019}
\Name{Mietke A., Jemseena V., Kumar K.~V., Sbalzarini I.~F. \and Juelicher F.}
  \REVIEW{Phys. Rev. Lett.}{123}{2019}{{188101}}.

\bibitem{Nitschke_PRSA_2020}
\Name{Nitschke I., Reuther S. \and Voigt A.} \REVIEW{Proc. Roy. Soc.
  A}{476}{2020}{20200313}.

\bibitem{Napoli_PRL_2012}
\Name{Napoli G. \and Vergori L.} \REVIEW{Phys. Rev. Lett.}{108}{2012}{207803}.

\bibitem{Nestler_SM_2020}
\Name{Nestler M., Nitschke I., L\"owen H. \and Voigt A.} \REVIEW{Soft
  Matter}{16}{2020}{4032}.

\bibitem{Nestler_arXiv_2021}
\Name{Nestler M. \and Voigt A.} \REVIEW{arXiv}{}{2021}{2107.07779}.

\bibitem{Nagai_PMB_2001}
\Name{Nagai T. \and Honda H.} \REVIEW{Phil. Mag. B}{81}{2001}{699}.

\bibitem{Staple_EPJE_2010}
\Name{Staple D., Farhadifar R., Roeper J.-C., Aigouy B., Eaton S. \and
  Juelicher F.} \REVIEW{Europ. Phys. J. E}{33}{2010}{117}.

\bibitem{Fletcher_BPJ_2014}
\Name{Fletcher A., Osterfield M., Baker R. \and Shvartsman S.} \REVIEW{Biophys.
  J.}{106}{2014}{2291}.

\bibitem{Li_BPJ_2014}
\Name{Li B. \and Sun S.} \REVIEW{Biophys. J.}{10}{2014}{1532}.

\bibitem{Bi_PRX_2016}
\Name{Bi D., Yang X., Marchetti M. \and Manning M.} \REVIEW{Phys. Rev.
  X}{6}{2016}{021011}.

\bibitem{Sussman_PRR_2020}
\Name{Sussman D.~M.} \REVIEW{Phys. Rev. Res.}{2}{2020}{023417}.

\bibitem{Nonomura_PLOS_2012}
\Name{Nonomura M.} \REVIEW{PLoS ONE}{7}{2012}{e33501}.

\bibitem{Camley_PNAS_2014}
\Name{Camley B., Zhang Y., Zhao Y., Li B., Ben-Jacob E., Levine H. \and Rappel
  W.-J.} \REVIEW{Proc. Nat. Acad. Sci. (USA)}{111}{2014}{14770}.

\bibitem{Loeber_SR_2015}
\Name{Loeber J., Ziebert F. \and Aranson I.} \REVIEW{Sci. Rep.}{5}{2015}{9172}.

\bibitem{Palmieri_SR_2015}
\Name{Palmieri B., Bresler Y., Wirtz D. \and Grant M.} \REVIEW{Sci.
  Rep.}{5}{2015}{{11745}}.

\bibitem{Mueller_PRL_2019}
\Name{Mueller R., Yeomans J. \and Doostmohammadi A.} \REVIEW{Phys. Rev.
  Lett.}{122}{2019}{048004}.

\bibitem{Wenzel_JCP_2019}
\Name{Wenzel D., Praetorius S. \and Voigt A.} \REVIEW{J. Chem.
  Phys.}{150}{2019}{164108}.

\bibitem{Loewe_PRL_2020}
\Name{Loewe B., Chiang M., Marenduzzo D. \and Marchetti M.} \REVIEW{Phys. Rev.
  Lett.}{125}{2020}{038003}.

\bibitem{Wenzel_arXiv_2021}
\Name{Wenzel D. \and Voigt A.} \REVIEW{Phys. Rev. E}{184}{2021}{054410}.

\bibitem{Peyret_BJ_2019}
\Name{Peyret G., Mueller R., d'Alessandro J., Begnaud S., Marcq P., Mege R.-M.,
  Yeomans J., Doostmohammadi A. \and Ladoux B.} \REVIEW{Biophys.
  J.}{117}{2019}{464}.

\bibitem{balasubramaniamEtAl2021}
\Name{Balasubramaniam L., Doostmohammadi A., Saw T., Narayana G., Mueller R.,
  Dang T., Thomas M., Gupta S., Sonam S., Yap A., Toyama Y., M\'ege R.-M.,
  Yeomans J. \and Ladoux B.} \REVIEW{Nature Materials}{20}{2021}{1156}.

\bibitem{Marth_JRSI_2015}
\Name{Marth W., Praetorius S. \and Voigt A.} \REVIEW{J. Roy. Soc.
  Interf.}{12}{2015}{20150161}.

\bibitem{Marth_IF_2016}
\Name{Marth W. \and Voigt A.} \REVIEW{Interf. Focus}{6}{2016}{20160037}.

\bibitem{Wenzel_CMAM_2021}
\Name{Wenzel D., Nestler M., Reuther S., Simon M. \and Voigt A.}
  \REVIEW{Comput. Meth. Appl. Math.}{21}{2021}{683}.

\bibitem{Jain_arXiv_2021}
\Name{Jain H., Wenzel D. \and Voigt A.} \REVIEW{arXiv}{}{2021}{2108.04743}.

\bibitem{Fily_PRL_2012}
\Name{Fily Y. \and Marchetti M.} \REVIEW{Phys. Rev. Lett.}{108}{2012}{235702}.

\bibitem{Redner_PRE_2013}
\Name{Redner G., Baskaran A. \and Hagan M.} \REVIEW{Phys. Rev.
  E}{88}{2013}{012305}.

\bibitem{Wysocki_EPL_2014}
\Name{Wysocki A., Winkler R. \and Gompper G.} \REVIEW{EPL}{105}{2014}{48004}.

\bibitem{dziuk2013finite}
\Name{Dziuk G. \and Elliott C.~M.} \REVIEW{Acta Numerica}{22}{2013}{289}.

\bibitem{nestler2019finite}
\Name{Nestler M., Nitschke I. \and Voigt A.} \REVIEW{J. Comput.
  Phys.}{389}{2019}{48}.

\bibitem{Vey_CVS_2007}
\Name{Vey S. \and Voigt A.} \REVIEW{Comput. Vis. Sci.}{10}{2007}{57}.

\bibitem{Witkowski_ACM_2015}
\Name{Witkowski T., Ling S., Praetorius S. \and Voigt A.} \REVIEW{Adv. Comput.
  Math.}{41}{2015}{1145}.

\bibitem{Salvalaglio_MMAS_2021}
\Name{Salvalaglio M., Voigt A. \and Wise S.~M.} \REVIEW{Math. Meth. Appl.
  Sci.}{44}{2021}{5385}.

\bibitem{Backofen_IJNAM_2019}
\Name{Backofen R., Wise S.~M., Salvalaglio M. \and Voigt A.} \REVIEW{Int. J.
  Math. Anal. Mod.}{16}{2019}{192}.

\bibitem{Praetorius_NIC_2017}
\Name{Praetorius S. \and Voigt A.} \Book{Collective cell behavior - a
  cell-based parallelization approach for a phase field active polar gel model}
  in proc. of \Book{NIC Symposium 2018}, edited by \Name{Binder K., M\"uller M.
  \and Trautmann A.} 2018 pp. 369--376.

\bibitem{Thomson_PM_1904}
\Name{Thomson J.~J.} \REVIEW{Philos. Mag.}{7}{1904}{237}.

\bibitem{Backofen_PRE_2010}
\Name{Backofen R., Voigt A. \and Witkowski T.} \REVIEW{Phys. Rev.
  E}{81}{2010}{025701}.

\bibitem{Backofen_MMS_2011}
\Name{Backofen R., Graef M., Potts D., Praetorius S., Voigt A. \and Witkowski
  T.} \REVIEW{Multisc. Mod. Sim.}{9}{2011}{314}.

\bibitem{Praetorius_PRE_2018}
\Name{Praetorius S., Voigt A., Wittkowski R. \and L\"owen H.} \REVIEW{Phys.
  Rev. E}{97}{2018}{052615}.

\bibitem{Atia_NP_2018}
\Name{Atia L., Bi D., Sharma Y., Mitchel J.~A., Gweon B., Koehler S.~A., DeCamp
  S.~J., Lan B., Kim J.~H., Hirsch R., Pegoraro A.~F., Lee K.~H., Starr J.~R.,
  Weitz D.~A., Martin A.~C., Park J.~A., Butler J.~P. \and Fredberg J.~J.}
  \REVIEW{Nature physics}{14}{2018}{613}.

\bibitem{Hesselink_PV_1994}
\Name{Hesselink L., Levy Y. \and Lavin Y.} \REVIEW{Proceedings Visualization
  '94}{}{1994}{140}.

\bibitem{Allan_2021}
\Name{Allan D., Caswell T., Keim N., van~der Wel C. \and Verweij R.}
  \Book{{soft-matter/trackpy:Trackpy v0.5.0}} ({2021}).
\newline\url{{https://doi.org/10.5281/zenodo.4682814}}

\bibitem{Crocker_JoCaIS_1996}
\Name{Crocker J.~C. \and Grier D.~G.} \REVIEW{Journal of Colloid and Interface
  Science}{179}{1996}{298}.

\bibitem{ZhangDesernoTu_2020}
\Name{Zhang Y.-H., Deserno M. \and Tu Z.-C.} \REVIEW{Phys. Rev.
  E}{102}{2020}{012607}.

\bibitem{NestlerVoigt_2022}
\Name{Nestler M. \and Voigt A.} \REVIEW{arXiv}{}{2021}{2107.07779}.

\bibitem{Keberetal_2014}
\Name{Keber F., Loiseau E., Sanchez T., DeCamp S., Giomi L., Bowick M.,
  Marchetti M., Dogic Z. \and Bausch A.} \REVIEW{Science}{345}{2014}{1135}.

\bibitem{Alaimoetal_2017}
\Name{Alaimo F., K{\"o}hler C. \and Voigt A.} \REVIEW{Scientific
  Reports}{7}{2017}{}.

\end{thebibliography}
\bibliographystyle{eplbib}

\end{document}